\documentclass[runningheads,a4paper]{llncs} %
\usepackage{times} %
\usepackage{amssymb} %
\usepackage{graphicx} %
\usepackage{url} %
\usepackage[utf8]{inputenc}
\usepackage{graphicx}
\usepackage{amsmath}
\usepackage{multirow}
\usepackage[table,xcdraw]{xcolor}
\newcommand{\keywords}[1]{\par\addvspace\baselineskip
\noindent\keywordname\enspace\ignorespaces#1}

\begin{document}

% \title{Experiment 2 Documentation}
% \author{Brendan O'Connor}
% \date{June 2021}
\mainmatter  % start of an individual contribution

\title{Zero-shot Singing Technique Conversion}

\titlerunning{Zero-shot Singing Technique Conversion}

\author{Brendan O'Connor\inst{1}\and Simon Dixon\inst{1}\and George Fazekas\inst{1} \thanks{This research is funded by the EPSRC and AHRC Centre for Doctoral Training in Media and Arts Technology (EP/L01632X/1).}}
%
% if the names of the authors are too long for the running head, please use the format: AuthorA et al.
\authorrunning{O'Connor et al.}

\institute{Centre for Digital Music, Queen Mary University of London, UK\\ \email{b.d.oconnor@qmul.ac.uk}}


\maketitle

% \tableofcontents %remove this for conference
% \pagebreak %remove this for conference

%All links must be hyper links using /url - test this

\begin{abstract}
    %should contain at least 70 and at most 150 words
    In this paper we propose modifications to the neural network framework, AutoVC \cite{qian2019} for the task of singing technique conversion. This includes utilising a pretrained singing technique encoder which extracts technique information, upon which a decoder is conditioned during training. By swapping out a source singer's technique information for that of the target's during conversion, the input spectrogram is reconstructed with the target's technique. We document the beneficial effects of omitting the latent loss, the importance of sequential training, and our process for fine-tuning the bottleneck. We also conducted a listening study where participants rate the specificity of technique-converted voices as well as their naturalness. From this we are able to conclude how effective the technique conversions are and how different conditions affect them, while assessing the model's ability to reconstruct its input data.

    
\end{abstract}

\section{Introduction}

Voice conversion (VC) is the task of converting the timbre of the voice so that the linguistic content is perceived to be spoken by a different person. It has been explored in relation to both singing and speech, which both possess different attributes consideration. Singing voice analysis is considerably more focused on sustained notes, harmonic/rhythmic structure, and relative pitch. In speech, these musical values are non-existent. Instead there is greater emphasis on aperiodic aspects, such as consonant utterances and rapidly shifting spectral envelopes. Tasks like VC and text-to-speech are in far more demand in the industry than singing-related tasks, and have therefore monopolised the spotlight in voice analysis and synthesis research. The latest approaches towards VC achieving state-of-the-art conversions utilise probabilistic machine learning techniques. Public domain speech datasets also vastly overshadow singing datasets in size and availability \cite{meseguer-brocal2020}, and so there is still much to be explored in relation to singing analysis and synthesis.

In this paper we tackle the task of singing technique conversion (STC) - the task of converting a singing technique without affecting the perceived identity of the singer, musical structure or linguistic content. We define singing technique as the method of voice production to achieve different timbres by adjusting the airflow, vocal folds, vocal tract shape, and sympathetic vibrations in the body \cite{heidemann2016}. We regard STC as a variation of voice conversion (VC), where the possibilities of voice transformation are restricted to be within a realistic variance of timbre for any given singer. We chose the term singing \textit{techniques} as opposed to singing style, due to the latter term's inconsistent use in literature, often referring to a range of very different audio and musical attributes due to its lack of reference to a concrete audio or singing concept.

To achieve STC, we apply a neural network model in the form of the conditioned autoencoder, AutoVC \cite{qian2019}. We discuss certain adaptations made to the architecture and investigate the effects of training it on different permutations of several datasets. To evaluate the model's ability to perform STC, we had participants rate the naturalness of the voice and guess what the target singing technique was supposed to be. Examples of audio used in this listening test can be found online.\footnote{https://github.com/Trebolium/singing\_technique\_conversion}

Real-time pitch correction algorithms have become commonplace in the music industry and influence the characteristics of modern pop singers today. We believe that the refined task of STC could have a similar influence on music production as it opens up the possibility of artistically manipulating a singer's \textit{performance}, rather than just quantising their pitch.  Over the last 5 years, many machine-learning approaches have been proposed to tackle voice transformation for speech (as discussed in the next section), but much less attention has been given to transforming the expression of the singing voice.

\section{Related Work} \label{Sec:relWork}
 
%Disentanglement is the process of separating attributes from data, and can allow us to do interesting manipulations and analysis with that data, or reassamble them in different combinations.

% Generalising disentanglement
Recent research in VC has been based on neural networks, which have influenced the frameworks proposed in this paper. \cite{jia2019} conditioned an autoencoder (trained on linguistic data) on speaker embeddings generated from a separately trained classifier network. During inference, these embeddings could be replaced to achieve VC. AutoVC \cite{qian2019} adapted this method to work with spectrograms, which will be described in detail in Section \ref{sec:training}. This was improved upon by conditioning the network on pitch contours to enforce prosody during conversion \cite{qian2020}, and further disentanglement was achieved for timbre, pitch contours, rhythm and utterances simultaneously by utilising 3 separate bottlenecks with different restrictions \cite{qian2020a}.  \cite{wu2020} achieve VC by using vector quantisation to separate speaker and content information, and later utilised U-nets \cite{ronneberger2015} to compensate for information lost during vector quantisation.

% VAE for VC
The application of the variational autoencoder (VAE) is well suited for `many-to-many' conversions (where all examples used for inference are seen during training). \cite{kameoka2020} use fully convolutional VAEs, conditioned on acoustic features, to perform VC. They combine spectral features of both converted and unconverted reconstructed audio in order to avoid over-smoothing - a known issue with VAEs. While VAEs present an elegant framework, they produce `blurry' results. Generative Adversarial Networks (GANs) have been known to reproduce better quality reconstructions of images than VAEs. However as they come without an autoencoder they are harder to train and suffer from `mode collapse', and there has not yet been an elegant proposal for combining VAEs with GANs \cite{tolstikhin2019}. The use of VAEs has the added benefit of utilising unsupervised learning, which bypasses the issue of low resources regarding labelled singing datasets. \cite{hsu2019} used Gaussian-Mixture VAEs (GMVAEs) for controllable speech synthesis, modelling the different attributes of speech as separate prior distributions before combining them in a VAE.
%AutoSVC - singing
For singing voice conversion, \cite{nercessian2020} adapted AutoVC by conditioning the network on pitch contours transposed to a suitable register for the converted singing, achievable through the implementation of a vocoder. 
%GANs - singing
\cite{lee2019} utilised a Wasserstein-GAN framework, using a decoder for pitch contours and another for generating `formant masks'. The product of these two decoders is the estimated mel-spectrogram for singing. They later explored the capabilities of this framework to achieve timbre and singing style disentanglement \cite{lee2020}, where a \textit{singing query} is converted into a singer identity embedding and used to condition both the pitch skeleton and formant-mask encoders on pitch modulation style and singer timbre, respectively.
%STC
\cite{luo2019} present the only other research we know of that addresses STC. They use GMVAEs to model singer and technique information to perform many-to-many conversions using a VAE architecture that utilises a convolutional recurrent neural network (CRNN) architecture.

The issue remains however, of what can be done with singing datasets which are small and few. \cite{nercessian2020} notes that the generalisation of the AutoVC framework allows it to be utilised as a Universal Background Model. \cite{basak2021a} synthesise monophonic singing datasets by superimposing pitch contours on existing speech datasets. \cite{chandna2020} use several autoencoder instances, trained separately on vocoder spectral data and music mixtures, while being conditioned on shared content embeddings and 1-hot speaker embeddings to produce a final network that is singer-independent and generates monophonic singing from musical mixtures. \cite{nachmani2019} generate novel speaker embeddings by combining embeddings from existing singers as a method of data augmentation. % These examples highlight some of the many innovative ways we can augment datasets and utilise datasets of a similar domain. There has also been a number of research citeXYZ\cite{bonada} that focuses on singing voice manipulation.
\section{Architecture}

%Why have we chosen conditioned AE?
%Needs to be stronger with more citations

% \cite{qian2019} says``neither GAN nor CVAE is perfect. GAN comes with a nice theoretical justification that the generated data would match the distribution of the true data, and has achieved state-of-the-art results, particularly in computer vision. However, it is widely acknowledged that GAN is very hard to train, and its convergence property is fragile. Also, although there is an increasing number of works that introduce GAN to speech generation (Donahue et al., 2018) and speech domain transfer (Pascual et al., 2017; Subakan and Smaragdis, 2018; Fan et al., 2018; Hosseini-Asl et al., 2018), there is no strong evidence that the generated speech sounds real. Speech that is able to fool the discriminators has yet to fool human ears''.

%\subsection{The AutoVC Framework} \label{sec:autovcFramework}
We use the AutoVC framework \cite{qian2019} for singing technique transformation, due to its elegant method of applying disentanglement. It is also capable of converting between source and target examples that have not been seen in the training datasets (zero-shot conversion). In AutoVC, a standard autoencoder architecture is conditioned on speaker embeddings that uniquely describe the timbre of a speaker to perform VC on spectrograms. These embeddings are generated by a pretrained speaker verification network \cite{wan2018}. The spectrograms are concatenated with these speaker embeddings, and fed through an encoder $E_{c}$, after which the encoded information is again concatenated with speaker embeddings before being fed to the decoder $D_{c}$. This conditioning, combined with careful calibration of an appropriate bottleneck size, allows the autoencoder to disentangle speaker timbre from utterance information. AutoVC also contains a `postnet' convolutional layer which is appended to the decoder to further develop a refined spectrogram from the decoder's output. After training, the speaker embeddings concatenated at the bottleneck can be swapped out to achieve VC. The loss function for AutoVC is a weighted combination of the self-reconstruction loss for both the decoder ($L_\mathit{decoder}$) and the postnet ($L_\mathit{postnet}$) output spectrograms, and the latent loss ($L_\mathit{latent}$). The latent loss represents the difference between the bottleneck's embedding $E_{c}(x)$ for the input x and its reconstructed form $E_{c}(\hat{x})$. This is summarised in Equation \ref{Eq:totalLoss}, where $\mu$ and $\lambda$ are empirically determined weights. Further details of AutoVC's architecture are given by \cite{qian2019}, which we follow in our implementation except for several adjustments discussed in this section.

\begin{equation}
    L_\mathit{total} = L_\mathit{decoder} +\mu L_\mathit{postnet} + \lambda L_\mathit{latent} .
    \label{Eq:totalLoss}
\end{equation}

%\subsection{The Singing Technique Encoder}
We will herein refer to our implementation of AutoVC as AutoSTC to reflect its purpose of STC. To facilitate this, we developed our own singing technique encoder (STE) to replace the external speaker encoder that was used in the original implementation. The STE is initially trained as a classifier. It takes a mel spectrogram as input, which is split into chunks of 0.5 seconds. These are fed in parallel through a neural network consisting of four 2D-convolutional layers (each of which is followed by batch normalisation, ReLU activation and max-pooling), two dense layers, two BLSTMs, a simplified attention mechanism \cite{raffel2016}, two more dense layers and finally a classification layer. This architecture was adapted from the VAE used by \cite{luo2019} and influenced by \cite{choi2017}. This network is able to achieve 86\% accuracy when classifying singing techniques within a test set of VocalSet (detailed in Section \ref{sec:training}, while our implementation of a 1D convolutional network on the waveform data as described by \cite{wilkins2018} only scored 57\%. During conversion, the STE's embedding preceding the classification layer are used for concatenation and conditioning with AutoVC as described above in place of the external speaker encoder embeddings.

\section{Training and Inference} \label{sec:training}

The Vocalset dataset \cite{wilkins2018} used to train the STE consists of recordings of 20 singers performing several musical exercises with different singing techniques. We chose a subset containing the techniques \textit{belt, straight, vibrato, lip trill, vocal fry} and \textit{breathy}, trimming off excess files that appear in one class but not the other, to yield a balanced class subset of  1182 examples (roughly 8K seconds). As the dataset is so small we only partition it into training and test sets by 8:2.

As \cite{nercessian2020} showed that the sequence of training of different datasets is important, AutoSTC was trained using subsets taken from VocalSet, VCTK \cite{veaux2017} and the raw singer recordings from MedleyDB \cite{bittner2014} in various permutations. All data was sampled at 16kHz and transformed into 80-bin mel spectrograms. While being trained on one dataset, AutoSTC was simultaneously tested on test sets from all three datasets in between training iterations (the VocalSet test set was the same set omitted when training the STE). We recorded the number iterations and loss values for each dataset where the loss showed no further improvement into Table \ref{tab:losses}, and transferred the saved neural network parameters of a nearby checkpoint to the proceeding dataset training session in the sequence. We trained AutoSTC once for every permutation of the datasets. Table \ref{tab:losses} shows that the order in which datasets are fed to the network does have a considerable impact on its loss. The paths \texttt{Vc->Vs->Md} (spanning 750k training steps) and \texttt{Vc->Md->Vs} (500k steps) led to the lowest loss values for MedleyDb and VocalSet reconstruction respectively, and were used to train models that generated the examples used in our listening test (see Section \ref{sec:exp}).

\begin{table}
\centering
\caption{Shows losses and training iterations (in parentheses) for VocalSet (left) and MedleyDB (right) along alternative paths. The optimum training path is highlighted in bold from left to right. Training that leads to an increase in loss is indicated with a circumflex, at which point that path is abandoned. For space, the dataset names are shortened as follows: VCTK:Vc, VocalSet:Vs, MedleyDB:Md.}
\scalebox{0.87}{
\begin{tabular}{|l|l|l|l|l|l|}
\hline
\multicolumn{6}{|c|}{Loss-Iteration for Vs}                                                                                                                                                              \\ \hline
\multicolumn{1}{|c|}{}                              &                                         & Vs         & \cellcolor[HTML]{FFFFFF}0.0274(100k) & Md           & \textasciicircum{}                           \\ \cline{3-6} 
\multicolumn{1}{|c|}{\multirow{-2}{*}{\textbf{Vc}}} & \multirow{-2}{*}{\textbf{0.0653(300k)}} & \textbf{Md} & \textbf{0.0386(150k)}                & \textbf{Vs} & \textbf{0.0268(50k)} \\ \hline
                                                    &                                         & Vc         & \textasciicircum{}                   & Md           & -                                            \\ \cline{3-6} 
\multirow{-2}{*}{Vs}                                & \multirow{-2}{*}{0.0347(150k)}          & Md          & \textasciicircum{}                   & Vs          & -                                            \\ \hline
                                                    &                                         & Vs         & 0.0290(50k)                          & Vc          & \textasciicircum{}                           \\ \cline{3-6} 
\multirow{-2}{*}{Md}                                 & \multirow{-2}{*}{0.0500(200k)}          & Vc         & \textasciicircum{}                   & Vc          & -                                            \\ \hline
\end{tabular}

\begin{tabular}{|l|l|l|l|l|l|}
\hline
\multicolumn{6}{|c|}{Loss-Iteration for Md}                                                                                                                                                                     \\ \hline
\multicolumn{1}{|c|}{}                              &                                         & \textbf{Vs} & \cellcolor[HTML]{FFFFFF}\textbf{0.0474(150k)} & \textbf{Md} & \textbf{0.0265(100k)}                      \\ \cline{3-6} 
\multicolumn{1}{|c|}{\multirow{-2}{*}{\textbf{Vc}}} & \multirow{-2}{*}{\textbf{0.0479(500k)}} & Md           & 0.0295(150k)                                  & Vs         & \textasciicircum{} \\ \hline
                                                    &                                         & Vc          & 0.0474(100k)                                  & Md          & 0.0301(50k)        \\ \cline{3-6} 
\multirow{-2}{*}{Vs}                                & \multirow{-2}{*}{0.0562(150k)}          & Md           & 0.0370(100k)                                  & Vs         & \textasciicircum{}      \\ \hline
                                                    &                                         & Vs          & \textasciicircum{}                            & Vc         & -                                          \\ \cline{3-6} 
\multirow{-2}{*}{Md}                                 & \multirow{-2}{*}{0.0367(150k)}          & Vc          & \textasciicircum{}                            & Vc         & -                                          \\ \hline
\end{tabular}

}
\label{tab:losses}
\end{table}

% EXTRA INFO: medleydb, which was broken up into chunks >= 3 seconds. In total, there were 1873 of these detected across both V1 and V2 of medleydb, with a total duration of 13134s, meaning that if we are using 3 second audio clips, we have 4378 potential 3 second examples. a subset of VCTK, which consists of 20 voices, and 44455 clips in total, each one defaulting to about three seconds. As the dataset is grouped by singers, an entire epoch would only contain 20 voices. Therefore to get through the entire dataset of clips would require 44455/20 = 2222 epochs. Qian2019 represented iterations as 'steps' which were measured by a single training example. As they proposed 100k examples, this means the data was seen 2.24 times. So depending on what we're using determines how many steps we consider to be in an epoch. In the case of VocalSet, we could consider it to be double the sample set are they are twice the example size in length. Schulter and Grill consider this to be determinable by considering how much each song is visited.

% \cite{jia2019} compare their proposed network against a baseline version which optimises a fixed embedding for each speaker in the training set, which basically learns a lookup table of speaker embeddings. This implies a finite number of embeddings from which the network can learn to predict. Using averaged STEs is similar to this, and would certainly allow for faster training. However we cannot assume that unlabelled data fits into any particular average embedding

% \subsection{Loss}

We found the L1 loss between decoder/postnet self-reconstructions and the input encouraged better convergence over L2 loss. We also tested the impact of excluding the latent loss for 100K steps for both VC and STC tasks. Results showed that training without latent loss performs significantly better for both tasks. The loss for STC with latent loss was 0.0237 (and 0.0185 without loss), while spectrograms were blurry and the audio lacked microtonal variation or vibrato, leaving a `bubbliness' artefact in its absence. Vowels were also not consistently reproduced. These shortcomings however are less noticeable for speech than singing. This result is worth highlighting, as latent loss has been used consistently in frameworks of a similar nature \cite{nercessian2020,jia2019,qian2019}.

% George says:
% CD loss tries to get the input to map to the same space in the latent representation
% It is going to embed speech in a very similar way to LPC
% There is much more continuous gradual change in singing - this is why the bottleneck embeddings may be so small
% Could invent some kind of restraint - cycle-consistency that has a tolerance value to make sure that the bn vector is in the right ‘area’
% There are techniques in singing that are not recognised in speech framework, so the cd has issues rerepresenting this

% \begin{table}
% \centering
% \begin{tabular}{lllll}
% \cline{1-3}
% \multicolumn{1}{|l|}{Condition} & \multicolumn{1}{l|}{Speaker Embs} & \multicolumn{1}{l|}{Singing Tech Embs} &  &  \\ \cline{1-3}
% \multicolumn{1}{|l|}{with CC}   & \multicolumn{1}{l|}{0.0210}       & \multicolumn{1}{l|}{0.0237}            &  &  \\ \cline{1-3}
% \multicolumn{1}{|l|}{w/o CC}    & \multicolumn{1}{l|}{0.0190}       & \multicolumn{1}{l|}{0.0185}            &  &  \\ \cline{1-3}
%                                 &                                   &                                        &  & 
% \end{tabular}
% \end{table}

% \subsection{Bottleneck dimensions}

Further preliminary trials allowed us to fine-tune AutoSTC's bottleneck. We analysed the resynthesised audio and noted that the net size of the feature space was more indicative of audio quality than focusing on time/frequency axis fine-tuning separately. We estimated the threshold to be a downsampling factor of 16 for the time-axis, with each timestep containing 32 features. Lower dimensionality representations resulted in deterioration in the reconstructed audio in a very similar manner to that of the network with latent loss included.

\section{Experiment Design} \label{sec:exp}

To evaluate our proposed network's ability to perform STC, we conducted a listening study, where 19 participants evaluated the converted audio for specificity and naturalness under the different conditions of models, gender, and source/target techniques used. The first of the models (Vs1) was trained on VocalSet alone and converted Vocalset data. The second (Vs2) was trained using the optimum path for Vocalset presented in Table \ref{tab:losses} to also convert VocalSet data, while the third (M1) used the optimum path for MedleyDB to convert MedleyDB data. Converted spectrograms were resynthesised using a pretrained wavenet model provided by the author of the AutoVC paper\footnote{https://github.com/auspicious3000/autovc}. Each model produced 8 random examples per participant, while adhering to a balanced representation of both gender and subset (train/test) conditions. To evaluate naturalness, we asked participants to consider how synthetic/realistic the voice itself sounded, and rate them on a scale of 1 (very unnatural) to 5 (very natural). In a separate task to evaluate specificity, participants were given a reference recording featuring a converted singing technique along with 6 unlabelled candidate recordings from the same singer to choose from. These candidate recordings were randomly selected from the relevant dataset partition, so that they each featured 1 of the 6 potential target techniques assigned to the reference recording. Participants were asked to select one recording they thought featured a singing technique closest to that of the reference recording, or more if the answer seemed ambiguous. In the case where reference recordings were converted MedleyDB examples, no ground truth labels existed, and so a singer of the same gender was randomly chosen from Vocalset to represent the 6 candidate singing techniques instead. Each of these tasks was presented 24 times. 6 resynthesised recordings of unconverted audio were also evaluated for naturalness. The interface was built using the Web Audio Evaluation Tool \cite{jillings2015}.
\section{Results}

The Mean Opinion Score (MOS) for unconverted data was 3.75 ± 0.34, and is important to consider when analysing the results of the study. This highlights the fact that a considerable amount of perceived naturalness has already been lost during the wavenet resynthesis process, and that the MOS values for technique conversion should be considered with this in mind. It is the comparison between conditions that we are interested in.

To calculate the similarity score $S$ for each condition, we used the formula in Equation \ref{eq:simscore}, where $P_n$ is a binary vector reflecting a participant's true/false predictions (identifying whether each candidate technique was the same as what was presented in the reference audio) for the $n$th task, $C_n$ is a 1-hot vector reflecting the correct technique for the task, and $N$ is the total number of tasks in the given condition. The similarity score is an average count of correct predictions weighted by the reciprocal of the number of predictions made for the corresponding task. %Wrong answers do not contribute to the similarity score.

\begin{equation}
    S =\frac{1}{N}\sum_{n=1}^N
    %\begin{cases}
        \frac{P_n . C_n}{||P_n||_1} .
        %& \text{if 1} \in P\\
        %0,              & \text{otherwise}
    %\end{cases} 
    \label{eq:simscore}
\end{equation}

% Denominator is the total number of options the participant selected to be similar

\begin{figure}[t]
    \centering
    \includegraphics[width=0.9\textwidth]{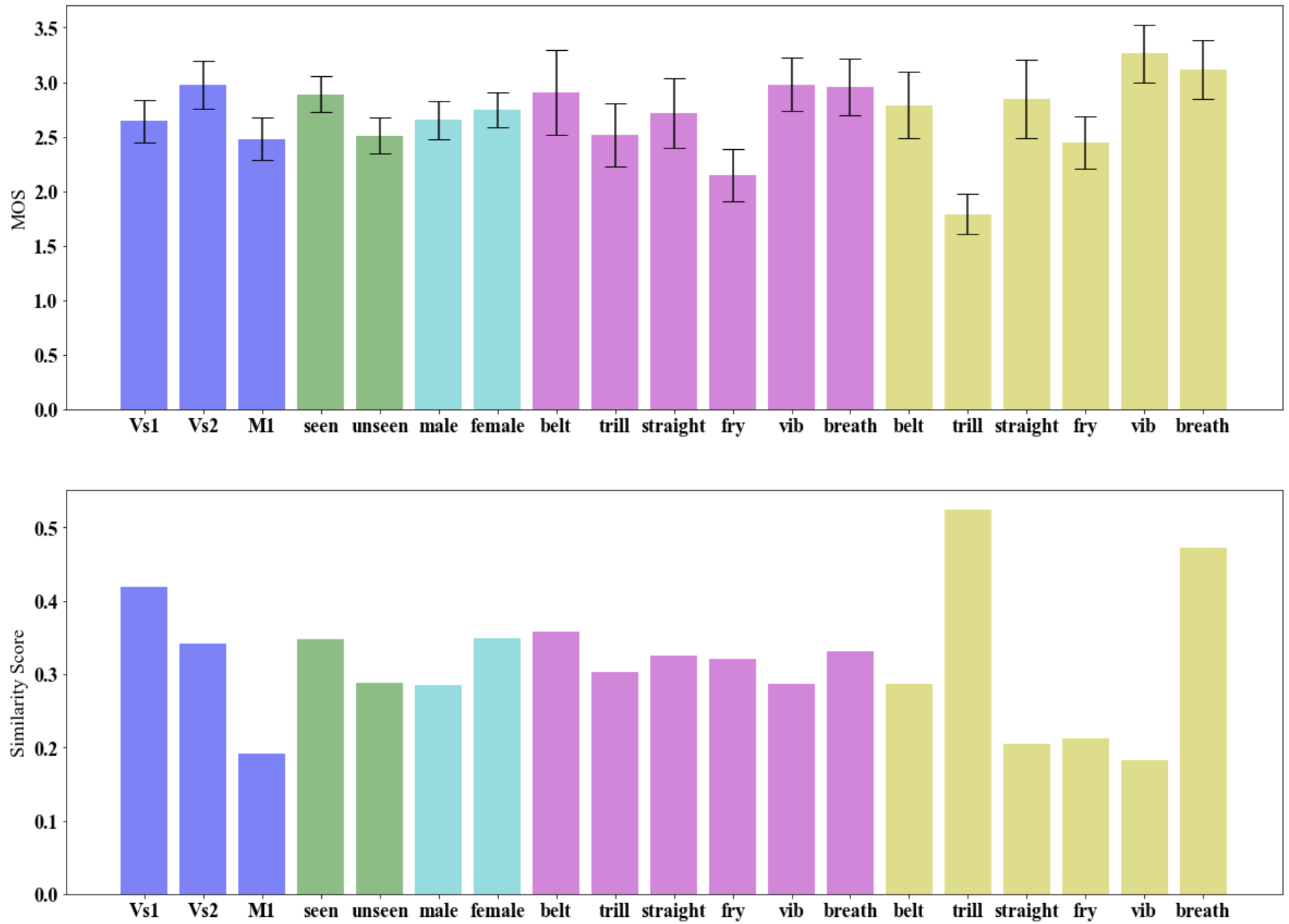}
    \caption{\textbf{Top:} Bar graph showing naturalness (MOS values and confidence intervals) for all conditions. The colours group together the conditions for (left to right): models, subsets, genders, source technique and target technique. \textbf{Bottom:} Bar graph showing similarity scores determined by Equation \ref{eq:simscore}, relative to correct answers.}
    \label{fig:expBarChart}
\end{figure}

Figure \ref{fig:expBarChart} displays the results obtained from the listening study.  The top graph displays MOS values for naturalness, with whiskers indicating the confidence intervals. The lower graph displays similarity scores. The combination of these two graphs give us insight into how each of our models perform, and what conditions influence the naturalness and specificity of the converted singing.

We detected from a Spearman's rank analysis that MOS and similarity scores were not significantly correlated. Similarity scores across all conditions measure higher than the chance level (0.16), which suggests that our models have some success in converting to the target techniques. The condition of source-technique groups does not significantly influence recognisability of the converted singing technique. However, the data would suggest that the features of the target techniques \textit{trill} and \textit{breathy} are significantly more distinguishable than the rest. \textit{Vibrato} scored the lowest for similarity, suggesting that this was a particularly difficult technique to synthesise convincingly. The reason for this is most likely due to the fact that VocalSet, upon which the STE network was trained, contains numerous examples labelled as \textit{belt} and \textit{straight}, while still featuring a considerable amount of frequency modulation (a unique feature of vibrato), making it difficult for AutoSTC to disentangle vibrato from other techniques effectively. It may also be the case that AutoSTC has difficulty disentangling vibrato from pitch contours. Alternatively it is possible that our models instead focused on altering the phonation modes associated with vibrato, which would be considerably less obvious to listeners than identifying whether frequency modulation is occurring.

The inclusion of all datasets in training Vs2 seemingly diminished its ability to accurately convert techniques (although the difference was not statistically significant). The M1 model scored significantly worse than the other models, which tells us that the features learned to generate technique embeddings from the STE network were not generalisable to data outside the dataset the VTE was trained on. There was also no statistically significant difference between gender and subset similarity scores.

In regards to MOS results, the target technique \textit{trill} scored lowest, suggesting that conversions to a trill technique may sound unnatural. Vs2 samples were significantly higher than Vs1 and M1, which suggests that providing the network with multiple datasets does improve its ability to synthesise natural sounding data. The target technique condition \textit{vibrato} scored the highest, but as mentioned above, this may be because the network is making changes more subtle than the frequency modulation which lessens the amount of transformation required, causing less synthetic artefacts. It is also perfectly possible that participants simply perceive the singing voice to be more natural when vibrato is present.

\section{Conclusion}

%VTE, 0-shot conversion multi-dataset training, good recon ≠ good specificity, not generalisable features, vibrato, vocalshit shit, unsupervised, vibrato-conversion
In this paper we have presented a network for vocal technique classification, and the first network to perform zero-shot conversion on singing techniques, achieving above chance level for all tested conditions. We have demonstrated that omitting latent loss and choosing the order in which AutoSTC was fed different datasets significantly diminished its reconstruction loss, improving its ability to reconstruct mel spectrograms. However we can conclude from the results of the listening study that this does not have any significant effect on AutoSTC's ability to perform technique conversion and may even diminish it. We therefore conclude that the features generated by supervised learning on the labelled VocalSet dataset are not sufficient to generalise to recordings of other singers. We also consider that the appearance of frequency modulation in other techniques in VocalSet may have forced the network to give less importance to this vibrato feature (we have however witnessed conversions where frequency modulation was synthesised, but in very limited cases, so we can not rule out the possibility that the AutoSTC framework is incapable of converting singing technique features beyond their spectral filter properties). The findings of our listening study are in agreement the vocal timbre maps generated in our previous research \cite{oconnor2020}.

Augmentation techniques such as those discussed in Section \ref{Sec:relWork} may improve the generalisation of the VTE to unseen data. We would also like to apply the Generalised End-to-End Loss techniques from \cite{wan2018} to the VTE and fine-tune its output embedding size. Due to shortcomings in labelled datasets, we will explore unsupervised/semi-supervised networks such as VAEs. It may also be worth investigating how AutoSTC performs when we condition it on further attributes such as speaker identity, pitch contours and vowel sounds. As we consider STC to be a restricted variation of VC and the fact that there are considerably larger datasets for speech, it may also be worth exploring the effects of pre-training an AutoVC framework for VC before switching its speaker encoder for the singing technique encoder and training it for STC. In future work we will also consider alternative options to the speech-trained wavenet vocoder as this has introduced artefacts to the audio that likely lowered MOS ratings for all audio. We have also observed that AutoSTC was unintentionally able to remove vibrato from singing when underfitting, which may be a capability worth fine-tuning in future work.

\end{document}